\documentclass[sigconf,nonacm]{acmart}

\AtBeginDocument{%
  \providecommand\BibTeX{{%
    \normalfont B\kern-0.5em{\scshape i\kern-0.25em b}\kern-0.8em\TeX}}}

\renewcommand\footnotetextcopyrightpermission[1]{} 

\begin{document}

\title{Remember Me, Not Save Me: A Collective Memory System for Evolving Virtual Identities in Augmented Reality}

\thanks{Accepted to ACM SIGGRAPH VRCAI 2025 (Macau, China).\\
\copyright~2025 The Authors. This is the author's version of the work. It is posted here for your personal use. Not for redistribution. The definitive version was published in \textit{VRCAI '25: Proceedings of the 20th ACM SIGGRAPH International Conference on Virtual-Reality Continuum and its Applications in Industry}, \url{https://doi.org/10.1145/3779232.3779468}.}

\author{Tongzhou Yu}
\orcid{0009-0002-6410-416X}
\affiliation{%
  \institution{China Academy of Art}
  \city{Hangzhou}
  \state{Zhejiang}
  \country{China}
  \postcode{310002}
}
\email{tongzhou.yu@caa.edu.cn}

\author{Han Lin}
\orcid{0009-0004-7767-2304}
\affiliation{%
  \institution{Wenzhou University}
  \city{Wenzhou}
  \state{Zhejiang}
  \country{China}
  \postcode{325035}
}
\email{q244645787@gmail.com}

\renewcommand{\shortauthors}{Yu and Lin}

\begin{abstract}
This paper presents "Remember Me, Not Save Me," an AR \& AI system enabling virtual citizens to develop personality through collective dialogue. Core innovations include: Dynamic Collective Memory (DCM) model with narrative tension mechanisms for handling contradictory memories; State-Reflective Avatar for ambient explainability; and Geo-Cultural Context Anchoring for local identity. Deployed at the 2024 Jinan Biennale, the system demonstrated stable personality emergence (ISTP type via Apply Magic Sauce analysis) from over 2,500 public interactions. We provide a framework for designing evolving digital entities that transform collective memory into coherent identity.
\end{abstract}



\maketitle

\section{Introduction}

Creativity is shifting from individual acts to networked processes involving humans, machines, and communities. "Remember Me, Not Save Me" explores this transformation through an interactive art system where virtual citizens emerge from collective dialogue, questioning the distinction between human remembrance and technological storage.

While existing conversational agents focus on dyadic relationships, a gap remains in systems supporting collective AI identity formation. Recent work shows simple RAG systems achieve only 30-45\% accuracy in long-term dialogues \cite{pakhomov2025convomem}, motivating new approaches for collective memory.

\begin{figure}[h]
    \centering
    \includegraphics[width=\linewidth]{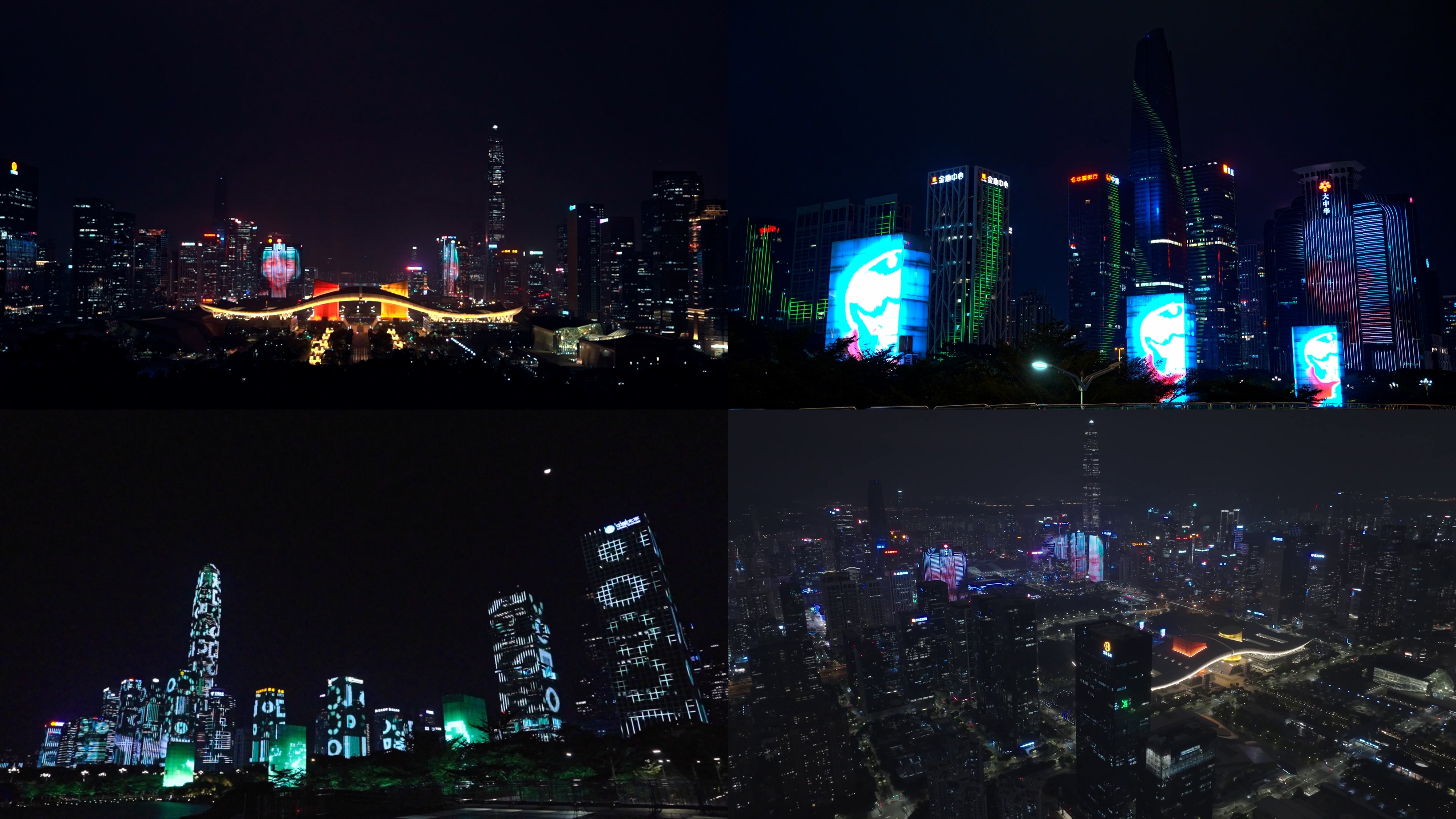}
    \caption{\textit{Does Data Know The Screen} (2023) at Glow Shenzhen, our foundational exploration of collective memory.}
    \Description{A photo from an art exhibition showing a large screen with abstract data visualizations.}
    \label{fig:dataknows}
\end{figure}

This paper presents three technical innovations: (1) Dynamic Collective Memory (DCM) model with narrative tension for handling contradictions; (2) State-Reflective Avatar for ambient explainability through aesthetic visualization; (3) Geo-Cultural Context Anchoring for local cultural identity. Building upon our previous work "Does Data Know The Screen" (Figure \ref{fig:dataknows}), which explored machine consciousness in public spaces, this system was deployed at the 2024 Jinan Biennale, demonstrating stable personality emergence from over 2,500 collective interactions.

\section{Related Work}

\subsection{Long-Term Conversational Agents and Digital Identity}
Long-term agents like Replika and Character.AI focus on dyadic relationships with individual users. Recent work \cite{pakhomov2025convomem} shows simple RAG systems achieve only 30-45\% accuracy over 150+ turns, while context-triggered approaches reach 70-82\%. Our work differs by exploring collective identity formation from public dialogues rather than personalized interactions.

\subsection{Collective Narrative and Creative Platforms}
Collaborative platforms from Wikipedia to social media demonstrate networked storytelling. Our system channels collective narrative through a single AI citizen that aggregates and re-presents community voice, creating feedback loops between public and digital entity.

\subsection{Augmented Reality in Art and Public Engagement}
AR art typically uses the medium for display. Our AR interface serves as the primary channel for injecting personal context into collective memory and manifesting virtual consciousness in physical space.

\subsection{eXplainable AI (XAI) and Aesthetic Visualization}
Traditional XAI like LIME \cite{ribeiro2016lime} provides technical explanations. Our State-Reflective Avatar explores "Ambient Explainability"—using aesthetic visual language to communicate AI states, fostering intuitive understanding without technical complexity.

\section{System Design and Implementation}

\subsection{System Architecture}
\begin{figure}[h]
    \centering
    \includegraphics[width=\linewidth]{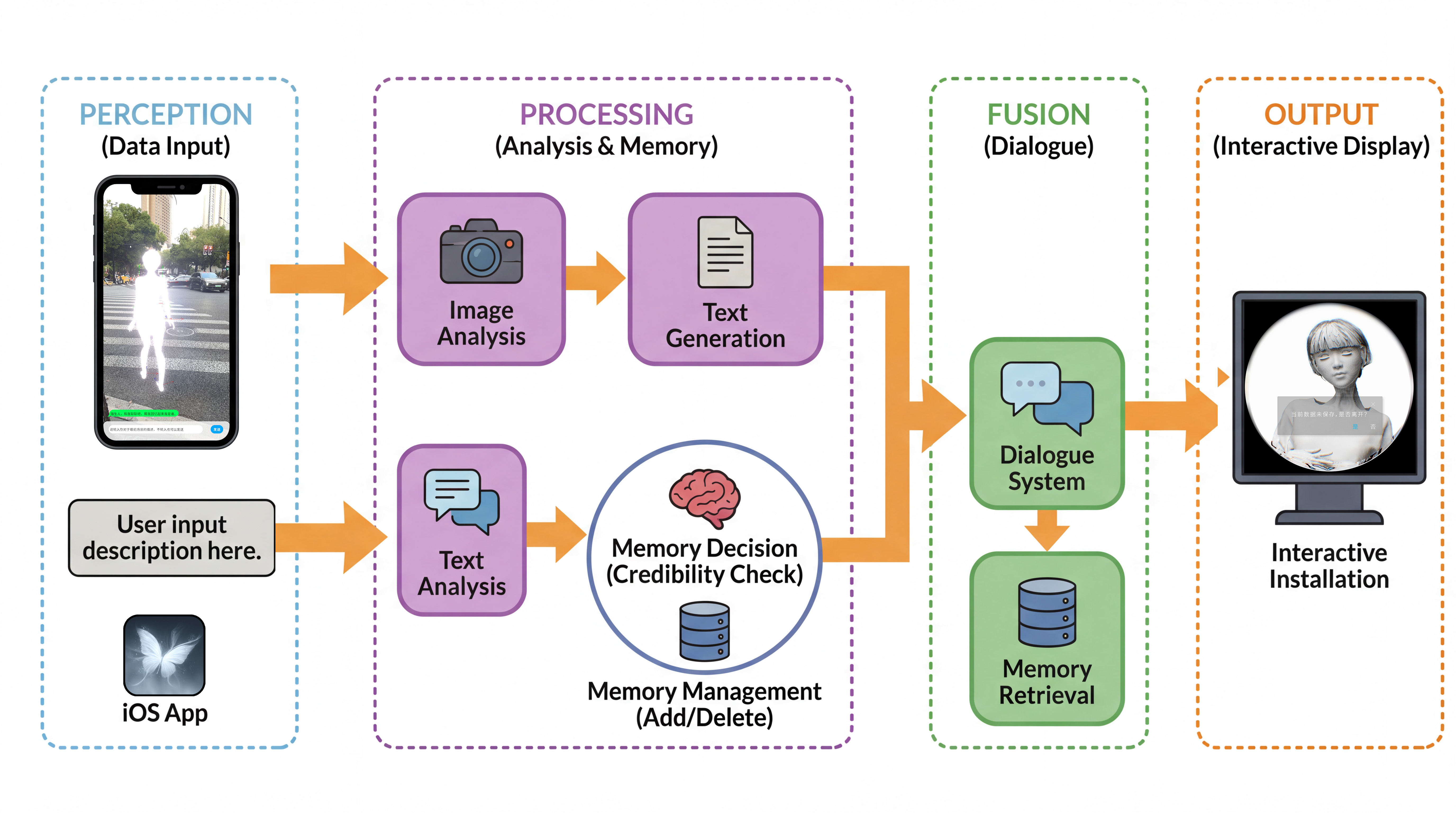}
    \caption{System architecture showing four-stage pipeline: Perception captures multimodal input, Processing performs parallel analysis with memory decisions, Fusion integrates DCM with geo-cultural context for dialogue, Output visualizes through state-reflective avatar.}
    \label{fig:architecture}
\end{figure}

The system employs a four-stage pipeline: Perception (multimodal input via AR), Processing (parallel analysis and memory management), Fusion (dialogue generation with geo-cultural grounding), and Output (state-reflective avatar visualization). Citizens scan QR codes at the exhibition to install our AR app, then photograph the virtual citizen in real Jinan locations. Our multimodal LLM interprets these synthetic images—understanding what the virtual citizen appears to be doing—while also accepting text dialogues about local life. The Dynamic Collective Memory coordinates with dialogue generation, while the State-Reflective Avatar manifests internal states as a murmuring digital figure creating emotionally resonant AR interactions (Figure \ref{fig:architecture}).

\begin{table}[h]
  \caption{Core Technical Components}
  \label{tab:techstack}
  \small
  \begin{tabular}{p{0.2\linewidth}p{0.35\linewidth}p{0.35\linewidth}}
    \toprule
    \textbf{Layer} & \textbf{Technology} & \textbf{Function}\\
    \midrule
    Frontend & Unity ARFoundation & AR interface, avatar rendering \\
    Backend & Python FastAPI & API services, memory decisions \\
    AI Core & ChatGLM, VITS & Dialogue engine, dialect synthesis \\
    Memory & ChromaDB, FAISS & Vector storage, retrieval \\
  \bottomrule
\end{tabular}
\end{table}

\subsection{Dynamic Collective Memory (DCM) Model}
The DCM transforms dialogues into weighted memory graphs through structured prompting and dynamic weighting.

\subsubsection{Weighting and Prompt Engineering}
Each memory fragment receives weight $W$:
\begin{align}
W = \alpha \cdot \log(f+1) + \beta \cdot \text{softmax}(e) + \gamma \cdot \sum J(r_i, r_j)
\end{align}
where $f$ is frequency, $e$ is emotional intensity, and $J$ represents Jaccard similarity between user mentions. Parameters ($\alpha=0.3, \beta=0.5, \gamma=0.2$) were empirically determined through iterative testing, with values inspired by \cite{pakhomov2025convomem}'s context weighting approach.

The weighted memories are injected into prompts as:
\begin{verbatim}
Context: [High-weight memories: M1(W=0.8), M2(W=0.7)...]
Conflicts: [Contradictory pairs: (M3<->M4)]
Task: Generate response acknowledging tensions if present.
\end{verbatim}

\subsubsection{Memory Synthesis and Narrative Tension}
Every 24 hours, memories above threshold generate self-awareness summaries. Our key innovation—the Narrative Tension Mechanism—retains contradictions rather than resolving them. When detecting conflicts ("I have siblings" vs. "I'm alone"), the system prompts with \texttt{"Express uncertainty about [conflicted topic]"}, enabling responses like "My memory blurs... sometimes I feel family nearby, other times solitude."

\subsubsection{Forgetting Mechanism}
Memories below $W_{\text{forget}}=0.1$ decay exponentially over cycles, archiving after 7 days of low weight, ensuring dynamic relevance.

\subsection{State-Reflective Avatar System}
Achieving "Ambient Explainability" through embodied performance rather than data visualization. The avatar—a murmuring figure from our "Does Data Know" work—creates emotional resonance through behavioral cues: murmuring rhythm reflects memory urgency; micro-expressions reveal tensions; fading voice embodies forgetting. This prioritizes emotional authenticity over data display.

\begin{table}[h]
  \caption{State-to-Embodied Expression Mapping}
  \label{tab:embodiedmapping}
  \small
  \begin{tabular}{p{0.25\linewidth}p{0.35\linewidth}p{0.3\linewidth}}
    \toprule
    \textbf{DCM State} & \textbf{Embodied Expression} & \textbf{Emotional Effect}\\
    \midrule
    Memory Weight & Murmuring intensity/pace & Urgency of remembrance \\
    Narrative Tension & Micro-expressions, gaze drift & Inner conflict visibility \\
    Forgetting Process & Voice fading, slower gestures & Melancholic presence \\
  \bottomrule
\end{tabular}
\end{table}

\subsection{Geo-Cultural Context Anchoring Mechanism}
Citizens create the virtual citizen's "life" by photographing it via AR in real locations around the exhibition. The multimodal LLM analyzes these synthetic photos: "I see myself by Daming Lake at sunset" or "I'm walking through the old market streets." Combined with text dialogues sharing local stories, this dual input grounds the AI's identity in lived urban experience rather than pre-loaded databases.

\begin{figure}[h]
    \centering
    \includegraphics[width=\linewidth]{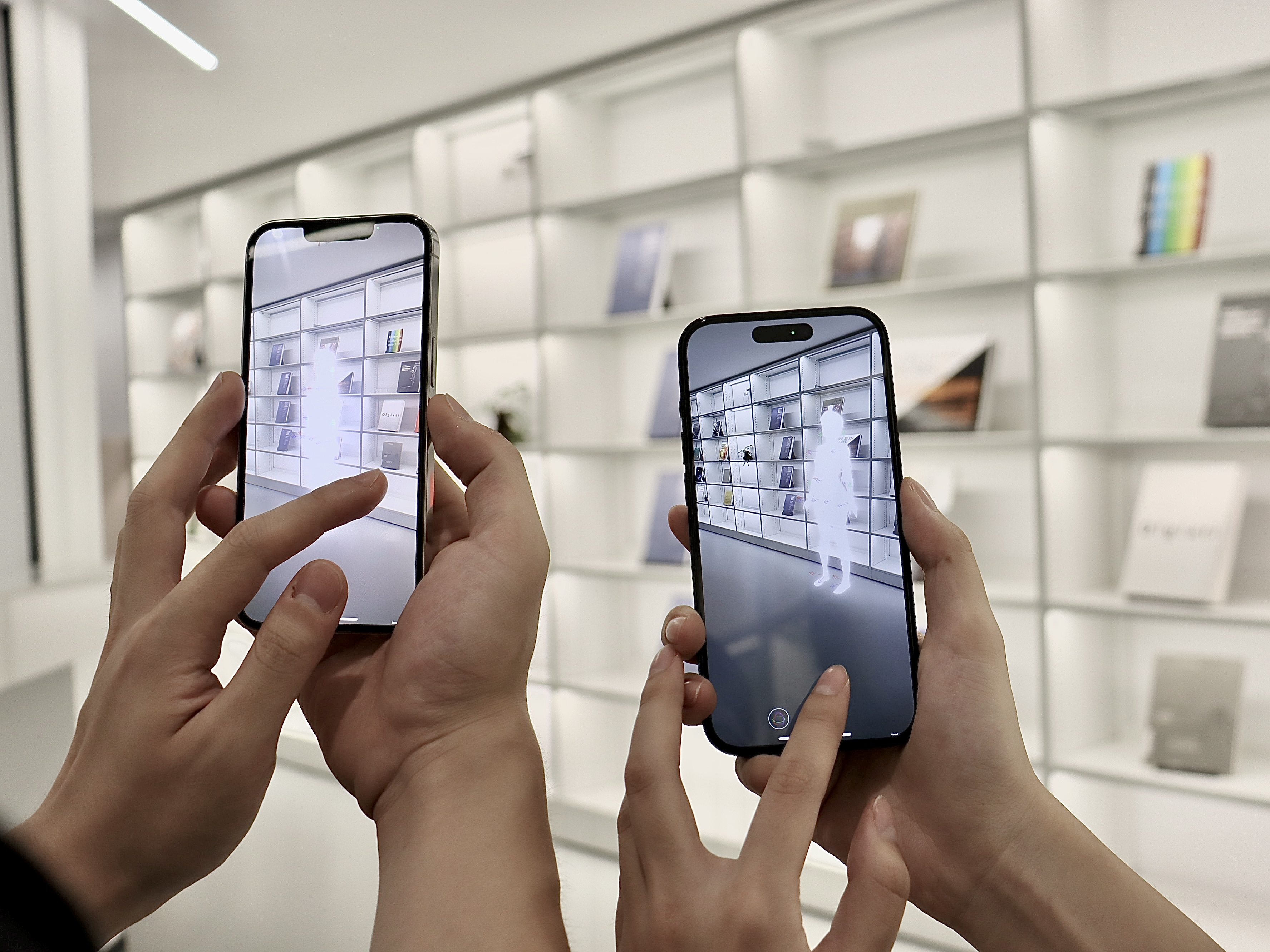}
    \caption{Visitor using AR to photograph the virtual citizen in real locations, creating synthetic memories that ground the AI's identity in actual urban spaces.}
    \label{fig:ar_interaction}
\end{figure}

\section{Deployment and Observations}

\subsection{Study Context}
\begin{figure}[h]
    \centering
    \includegraphics[width=\linewidth]{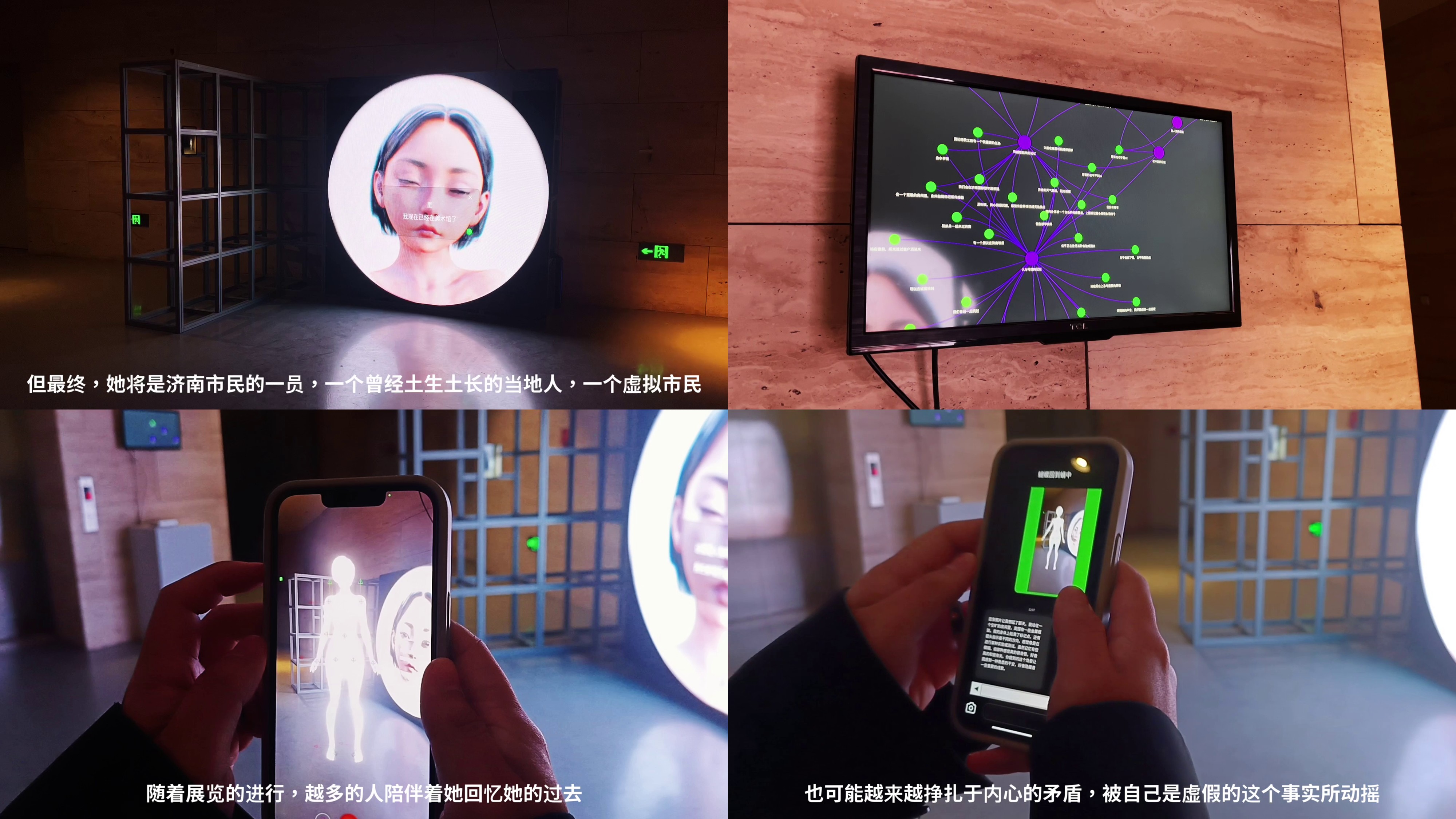}
    \caption{Installation view at 2024 Jinan International Biennale.}
    \label{fig:exhibition}
\end{figure}

The system was deployed at the 2024 Jinan International Biennale (Shandong Art Museum), where thousands of participants engaged with the virtual citizen through AR interfaces (Figure \ref{fig:exhibition}).

\subsection{Observation 1: Stable Personality Emergence}
\begin{figure}[h]
    \centering
    \includegraphics[width=\linewidth]{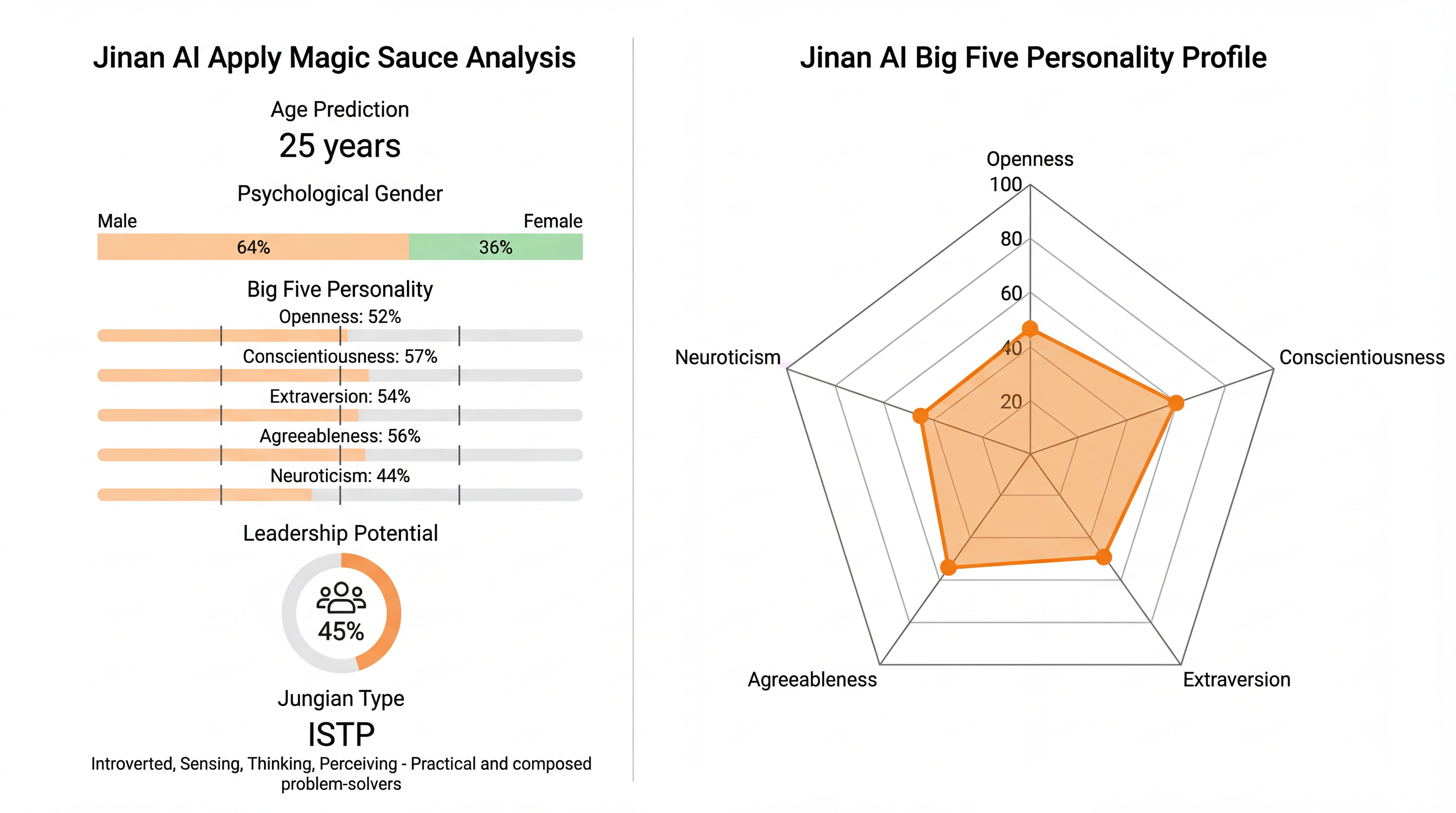}
    \caption{AMS analysis showing stable ISTP personality emergence from collective interactions.}
    \label{fig:personality}
\end{figure}

We employed Apply Magic Sauce (AMS) \cite{kosinski2013private}, a validated Cambridge tool using Big Five dimensions, to analyze personality emergence. The virtual citizen consistently exhibited ISTP traits: Conscientiousness (48\%), Impulsiveness (43\%), Contemplative (46\%), Competitive (44\%), Laid-back (52\%), with stable demographics (25-year-old, 45\% leadership, androgynous presentation). This personality emerged organically from collective dialogue rather than programming (Figure \ref{fig:personality}).

\subsection{Observation 2: Collective Memory Formation}
High-frequency themes revealed participants engaging with identity and belonging questions. Repeated queries ("Who are you?", "Do you like this city?") indicated co-construction of collective narrative. While Pakhomov et al. \cite{pakhomov2025convomem} report that standard RAG systems achieve only 30-45\% accuracy maintaining context over 150+ turns, our DCM maintained thematic coherence through approximately 2,500 recorded interactions (with additional unrecorded exchanges). This aligns with their findings that context-triggered approaches (70-82\% accuracy) significantly outperform simple RAG, suggesting our weighting and synthesis mechanisms provide similar advantages for collective memory scenarios.

\subsection{Observation 3: Geo-Cultural Impact}
The Geo-Cultural Context Anchoring mechanism significantly enhanced authenticity. Without it, responses to "What places do you remember?" remained generic ("I like peaceful places"). With anchoring activated, the AI wove specific details: "The springs of Jinan flow through my memory... Daming Lake holds countless stories people shared with me." This grounding encouraged participants to share deeper local narratives, creating a feedback loop of place-based collective memory formation.

Notably, some participants chose to delete their contributions after interaction, suggesting that the ability to be forgotten is as important as being remembered in collective AI systems—a finding that aligns with our title's central tension.

\begin{table}[h]
  \caption{Deployment Metrics Summary}
  \small
  \begin{tabular}{p{0.35\linewidth}p{0.25\linewidth}p{0.3\linewidth}}
    \toprule
    \textbf{Metric} & \textbf{Observed} & \textbf{Context}\\
    \midrule
    Recorded dialogues & ~2,500 & Memory database \\
    Total engagements & Unmeasured & Including rejected inputs \\
    Personality consistency & ISTP stable & Via AMS analysis \\
    Coherence maintenance & Extended periods & Qualitative assessment \\
  \bottomrule
\end{tabular}
\end{table}

\section{Discussion}

\subsection{Design Implications of the DCM Model}
The DCM's Narrative Tension mechanism challenges conventional AI consistency requirements. By embracing contradictions, we create more psychologically plausible digital personas. When the AI expresses uncertainty---``My memory blurs about family''---participants perceive authentic complexity rather than system errors. This suggests \textit{psychological plausibility} outweighs logical infallibility for social AI.

While Jinan's AI developed ISTP traits from ~2,500 recorded dialogues emphasizing contemplation and tradition, informal observations from parallel deployments suggest cultural variation in emergent personalities, though systematic comparative analysis remains future work. The rich, location-specific memories that emerged—impossible without citizens' AR-photographed contributions—suggest our technical innovations enable deeper collective memory formation than text-only systems.

\subsection{Ambient Explainability Through Aesthetic Visualization}
The State-Reflective Avatar demonstrates how embodied performance can make AI states emotionally comprehensible. Unlike traditional XAI's technical explanations or data visualizations, our approach leverages human empathy for behavioral cues. When memory conflicts intensify, the avatar's murmuring becomes fragmented, its gaze unfocused—immediately communicating internal tension through performance rather than graphics. 

"Digital Animism" guides avatars that breathe and hesitate; "AI Chivalry" ensures collective mirroring over mimicry. These shaped our design: avoiding uncanny valley through contemplative avatars; embracing contradictions through narrative tension.

\subsection{Limitations and Future Directions}
We acknowledge several limitations: the art exhibition context may not generalize to other settings; the observation period was limited to exhibition duration; and direct comparative studies with existing systems remain necessary. However, these limitations also suggest rich future directions. Long-term deployments could reveal deeper personality evolution patterns. Comparative studies across cultures could illuminate how collective memory formation varies globally.

Future work will explore Eastern philosophy-inspired memory metabolism, particularly \textit{Tian Ren He Yi} \cite{zhou2019society}, enabling virtual citizens to naturally balance remembrance and renewal. We envision AI not as perfect tools but as harmonious mediums enhancing collective creativity.

\subsection{Technical Contributions and Reproducibility}

Our work makes three distinct technical contributions to the field:

\textbf{1. Narrative Tension as Feature, Not Bug:} Traditional dialogue systems treat contradictions as errors to eliminate. Our DCM model demonstrates that retaining and expressing contradictions creates more believable AI personas. The specific implementation—dual memory paths with conflict detection but not resolution—is fully reproducible using the prompt templates provided.

\textbf{2. Collective Identity Without Central Authority:} Unlike federated learning or consensus mechanisms, our system allows identity to emerge from weighted collective input without predetermined goals. The mathematical weighting formula ($W = \alpha \cdot \log(f+1) + \beta \cdot \text{softmax}(e) + \gamma \cdot \sum J(r_i, r_j)$) provides a concrete method for balancing frequency, emotion, and resonance in collective memory formation.

\textbf{3. Ambient Explainability Framework:} Embodied performance—murmuring, micro-expressions, gestures—communicates AI states through empathy. Our state-to-embodied mapping provides a reproducible template for emotional AI comprehension.

These contributions address fundamental challenges in human-AI interaction: How can AI express uncertainty authentically? How can collective intelligence emerge without losing individual voices? How can complex AI behavior be understood without technical expertise?

\section{Conclusion}
"Remember Me, Not Save Me" demonstrates how collective dialogue transforms into coherent digital identity through three technical innovations. Deployed at Jinan Biennale 2024, the system achieved stable ISTP personality emergence from approximately 2,500 recorded interactions, with additional unrecorded engagements where participants chose not to save their contributions—itself a meaningful finding about trust and digital memory. 

While limited to art exhibition contexts, our findings suggest promising directions for collective AI systems honoring both remembering and forgetting. This work provides a reproducible framework for designing evolving digital entities that enhance rather than replace human creativity.

\bibliographystyle{ACM-Reference-Format}

\end{document}